\documentclass[%
 reprint,
 groupedaddress,
 showpacs, 
 noshowkeys,
 amsmath,amssymb,
 aps,
 prl,
]{revtex4-1}

\usepackage{graphicx}
\usepackage{dcolumn}
\usepackage{bm}

\usepackage[cdot,mediumqspace,amssymb]{SIunits} 
\usepackage[autolanguage]{numprint} 

\usepackage{ifthen}
\usepackage{ifpdf}
\ifpdf
    \graphicspath{{Figures/PDF/}}
\else
    \graphicspath{{Figures/EPS/}}
\fi

\usepackage{cleveref}

\usepackage{color}

\begin{document}

\title{Cross-Linked Structure of Network Evolution}

\author{Danielle S. Bassett$^{1,2,3,*}$, Nicholas F.
Wymbs$^{4}$, Mason A. Porter$^{5,6}$, Peter J. Mucha$^{7,8}$, Scott T.
Grafton$^{4}$} \affiliation{$^{1}$Department of Bioengineering, University of
Pennsylvania, Philadelphia, PA 19104, USA;\\ $^{2}$Department of Physics,
University of California, Santa Barbara, CA 93106, USA;\\ $^{3}$ Sage Center
for the Study of the Mind, University of California, Santa Barbara, CA 93106;\\
$^{4}$ Department of Psychology and UCSB Brain Imaging Center, University of California, Santa Barbara, CA 93106, USA;\\
$^{5}$ Oxford Centre for Industrial and Applied Mathematics, Mathematical
Institute, University of Oxford, Oxford OX1 3LB, UK;\\ $^{6}$ CABDyN
Complexity Centre, University of Oxford, Oxford, OX1 1HP, UK;\\
$^{7}$Carolina Center for Interdisciplinary Applied Mathematics, Department
of Mathematics, University of North Carolina, Chapel Hill, NC 27599, USA;\\
$^{8}$Department of Applied Physical Sciences, University of North Carolina, Chapel Hill, NC 27599, USA;\\
$^*$Corresponding author. Email address: dsb@seas.upenn.edu}

\date{\today}



\begin{abstract}

We study the temporal co-variation of network co-evolution via the
\emph{cross-link structure} of networks, for which we take advantage of the
formalism of hypergraphs to map cross-link structures back to network nodes.
We investigate two sets of temporal network data in detail. In a network of
coupled nonlinear oscillators, hyperedges that consist of network edges with
temporally co-varying weights uncover the driving co-evolution patterns of
edge weight dynamics both within and between oscillator communities. In the
human brain, networks that represent temporal changes in brain activity
during learning exhibit early co-evolution that then settles down with
practice, and subsequent decreases in hyperedge size are consistent with
emergence of an autonomous subgraph whose dynamics no longer depends on other
parts of the network. Our results on real and synthetic networks give a
poignant demonstration of the ability of cross-link structure to uncover
unexpected co-evolution attributes in both real and synthetic dynamical
systems.  This, in turn, illustrates the utility of analyzing cross-links for
investigating the structure of temporal networks.

\end{abstract}


\pacs{89.75.Fb, 89.75.Hc, 87.19.L-}

\keywords{Temporal Networks, Oscillators, Neuroscience, Hypergraphs}

\maketitle


\textbf{Networks provide a useful framework for gaining insights into a wide
variety of  social, physical, technological, and biological phenomena
\cite{Newman2010}. As time-resolved data become more widely available, it is
increasingly important to investigate not only static networks but also
temporal networks \cite{Holme2011,Holme2013}. It is thus critical to develop
methods to quantify and characterize dynamic properties of nodes (which
represent entities) and/or edges (which represent ties between entities) that
vary in time. In the present paper, we describe methods for the
identification of cross-link structures in temporal networks by isolating
sets of edges with similar temporal dynamics. We use the formalism of
hypergraphs to map these edge sets to network nodes, thereby describing the
complexity of interaction dynamics in system components. We illustrate our
methodology using temporal networks that we extracted from synthetic data
generated from coupled nonlinear oscillators and empirical data from human
brain activity.}


\section{Introduction}

Many complex systems can be represented as temporal networks, which consist
of components (i.e., nodes) that are connected by time-dependent edges
\cite{Holme2011,Holme2013}. The edges can appear, disappear, and change in
strength over time.  To obtain a deep understanding of real and model
networked systems, it is critical to try to determine the underlying drivers
of such edge dynamics. The formalism of temporal networks provides a means to
study dynamic phenomena in biological
\cite{Bassett2011b,Bassett2012b,Wymbs2012}, financial
\cite{Fenn2009,Fenn2011}, political \cite{waugh09,Mucha2010,Macon2012},
social \cite{Fararo1997,andrea2011,Onnela2007,Wu2010,morenotwitter,Christakis07,Snijders07}
systems, and more.

Capturing salient properties of temporal edge dynamics is critical for
characterizing, imitating, predicting, and manipulating system function.
Let's consider a system that consists of the same $N$ components for all
time. One can parsimoniously represent such a temporal network as a
collection of edge-weight time series. For undirected networks, we thus have
a total of $N(N-1)/2$ time series, which are of length $T$.  The time series
can either be inherently discrete or they can be obtained from a
discretization of continuous dynamics (e.g., from the output of a continuous
dynamical system). In some cases, the edge weights that represent the
connections are binary, but this is not true in general.

Several types of qualitative behavior can occur in time series that represent
\emph{edge dynamics} \cite{Slotine2012,Nepusz2012}. For example, unvarying
edge weights are indicative of a static system, and independently varying
edge weights indicate that a system does not exhibit meaningfully correlated
temporal dynamics.  A much more interesting case, however, occurs when there
are meaningful transient or long-memory dynamics.  As we illustrate in this
article, one can obtain interesting insights in such situations by examining
network \emph{cross-links}, which are defined via the temporal co-variation in
edge weights.  Illuminating the structure of cross-links has the potential to
enable predictability.

To gain intuition about the importance of analyzing cross-links, it is useful
to draw an analogy from biology. The cellular cytoskeleton \cite{Peters1963}
is composed of actin filaments that form bridges (edges) between different
parts (nodes) of a cell. Importantly, the bridges are themselves linked to
one another via actin-binding proteins. Because the network edges in this
system are not independent of each other, the structure of cross-links has
important implications for the mechanical and transport properties of the
cytoskeleton. Similarly, one can think of time-dependent relationships
between edge weights as cross-links that might change the temporal landscape
for dynamic phenomena like information processing, social adhesion, and
systemic risk. Analyzing cross-links allows one to directly investigate
time-dependent correlations in a system, and it thereby has the potential to
yield important insights on the (time-dependent) structural integrity of a
diverse variety of systems.

\begin{figure}
\begin{center}
\includegraphics[width=.95\linewidth]{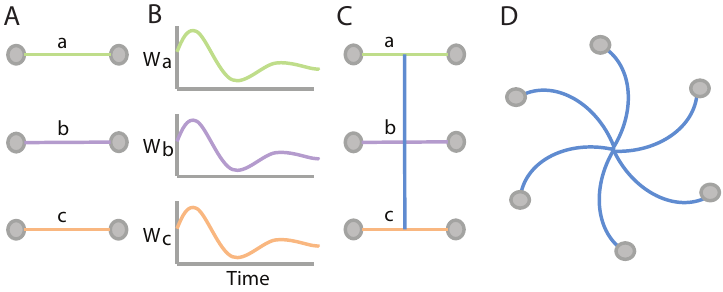}
\caption[]{\textbf{Co-Evolution Cross-Links and Hyperedges.} A set of
(\emph{A}) node-node edges with \emph{(B)} similar edge-weight time series
are  \emph{(C)} cross-linked to one another, which yields \emph{(D)} a
hyperedge that connects them.
    \label{FigHyperEdge}}
    \end{center}
\end{figure}

In this article, we develop a formalism for uncovering the structure in
time-dependent networks by extracting groups of edges that share similar
temporal dynamics. We map these cross-linked groups of edges back to the
nodes of the original network using hypergraphs \cite{Bollobas1998}. We
define a \emph{co-evolution hypergraph}\footnote{In this paper, we use the
term \emph{co-evolution} to indicate temporal co-variation of edge weights in
time.  The term co-evolution has also been used in other contexts in network
science (e.g., \cite{Xie2007,Kim2013}).} via a set of hyperedges that
captures cross-links between network edges, where each hyperedge is given by
the set of edges that exhibit statistically significant similarities to one
another in the edge-weight time series (see Fig.~\ref{FigHyperEdge}). A
single temporal network can contain multiple hyperedges, and each of these
can capture a different temporal pattern of edge-weight variation.

We illustrate our approach using ensembles of time-dependent networks
extracted from a nonlinear oscillator model and empirical neuroscience data.


\section{Cross-Link Structure}

To quantify network co-evolution, we extract sets of edges whose weights
co-vary in time. For a temporal network $\bf{A}_{t}$, where each $t$ indexes
a discrete sequence of $N\times N$ adjacency matrices, we calculate the $E
\times E$ adjacency matrix ${\bf \Lambda}$, where the matrix element
$\Lambda_{ab}$ is given by the Pearson correlation coefficient between the
time series of weights for edge $a$ and that for edge $b$. Note that
$E=N(N-1)/2$ is the total number of possible (undirected) edges per layer in
a temporal network.  The layers can come from several possible sources: data
can be inherently discrete, so that each layer represents connections at a
single point in time; the output of a continuous system can be discretized
(e.g., via constructing time windows); etc. We identify the statistically
significant elements of the edge-edge correlation matrix ${\bf \Lambda}$ (see
the Supplementary Material \footnote{Supplementary Material for this
manuscript can be found at [URL will be inserted by AIP]}), and we retain
these edges (with their original weights) in a new matrix ${\bf
\Lambda^{\prime}}$.  We set all other elements of ${\bf \Lambda^{\prime}}$ to
$0$.

We examine the structure of the edge-edge co-variation represented by the
$E\times E$ matrix ${\bf \Lambda^{\prime}}$ by identifying sets of edges that
are connected to one another by significant temporal correlations (i.e., by
identifying \emph{cross-links}; see Fig.~\ref{FigHyperEdge}). If ${\bf
\Lambda^{\prime}}$ contains multiple connected components, then we study each
component as a separate edge set. If ${\bf \Lambda^{\prime}}$ contains a
single connected component, then we extract edge sets using community
detection. (See the Supplementary Material [24] for a description of the
community-detection techniques that we applied to the edge-edge association
matrix.) We represent each edge set as a hyperedge, and we thereby construct
a co-evolution hypergraph $\bf{H}$. The nodes are the original $N$ nodes in
the temporal network, and they are connected via a total of $\eta$ hyperedges
that we identified from ${\bf \Lambda^{\prime}}$. The benefit of treating
edge communities as hyperedges is that one can then map edge communities back
to the original network nodes. This, in turn, makes it possible to capture
properties of edge-weight dynamics by calculating network diagnostics on
these nodes.


\textbf{Diagnostics.} To evaluate the structure of co-evolution hypergraphs,
we compute several diagnostics. To quantify the extent of co-evolution, we
define the \emph{strength} of co-variation as the sum of all elements in the
edge-edge correlation matrix: $\nu_{A_{t}} = \sum_{a,b}
\Lambda^{\prime}_{ab}$. To quantify the breadth of a single co-variation
profile, we define the \emph{size} of a hyperedge as the number of
cross-links that comprise the hyperedge: $s(h) = \frac{1}{2} \sum_{a,b \in
\lambda} [\Lambda^{\prime}_{ab}>0]_{\lambda}$, where the square brackets
denote a binary indicator function (i.e., 1 if is true and 0 if it is false)
and $\lambda$ indicates the set of edges that are present in the hyperedge
$h$ of the matrix ${\bf \Lambda^{\prime}}$. To quantify the prevalence of
hyperedges in a single node in the network, we define the \emph{hypergraph
degree} of a node $i$ to be equal to the number of hyperedges $\eta_{i}$
associated with node $i$.

\begin{figure}
\begin{center}
\includegraphics[width=0.95\linewidth]{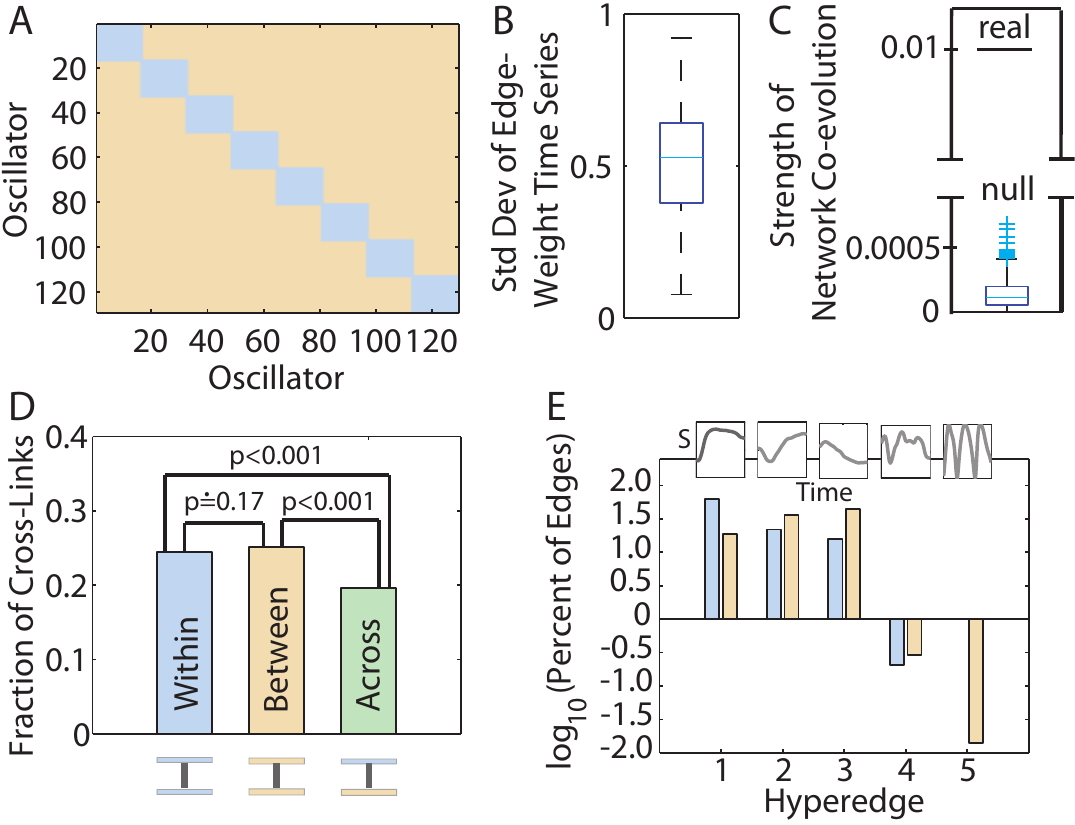}
\caption[]{\label{Fig:Kuramoto} \textbf{Co-evolution Properties of Kuramoto
Oscillator Network Dynamics.} \emph{(A)} Community structure in a network of
Kuramoto oscillators.  \emph{(B)} A box plot of the standard deviation in
edge weights over time for a temporal network of Kuramoto oscillators.
\emph{(C)} Strength of network co-evolution $\nu_{A_{t}}$ of the real
temporal network and a box plot indicating the distribution of $\nu_{A_{t}}$
obtained from 1000 instantiations of a null-model network. \emph{(D)}
Fraction of significant edge-edge correlations (i.e., \emph{cross-links})
that connect a pair of within-community edges (``\emph{Within}"), that
connect a pair of between-community edges (``\emph{Between}"), and that
connect a within-community edge to a between-community edge
(``\emph{Across}"). We calculated the statistical significance of differences
in these fraction values across the 3 cross-link types by permuting labels
uniformly at random between each type of pair. \emph{(E)} Fraction of (blue)
within-community and (peach) between-community edges in each of the 5 edge
sets extracted from ${\bf \Lambda^{\prime}}$ using community detection. We
give values on a logarithmic scale. \emph{Insets} Mean synchronization [$S(t)
= \sum_{(i,j) \in h} A_{ij}(t)$] of these edges as a function of time for
each hyperedge $h$.
    }
    \end{center}
\end{figure}


\section{Networks of Nonlinear Oscillators}

Synchronization provides an example of network co-evolution, as the coherence
(represented using edges) between many pairs of system components (nodes) can
increase in magnitude over time \cite{scholarsync,pikovsky}. Pairs of
edge-weight time series exhibit temporal co-variation (i.e., they have nontrivial
cross-links) because they experience such a trend. Perhaps less intuitively,
nontrivial network co-evolution can also occur even without synchronization.
To illustrate this phenomenon, we construct temporal networks from the
time-series output generated by interacting Kuramoto oscillators
\cite{Kuramoto84}, which are well-known dynamical systems that have been
studied for their synchronization properties (both with and without a
nontrivial underlying network structure)
\cite{strogatz00,scholarsync,pikovsky,Arenas2008,Shima2004,Abrams2004,Arenas2006,Stout2011}.
By coupling Kuramoto oscillators on a network with community structure
\cite{Arenas2006}, we can probe the co-evolution of edge weight time series
both within and between synchronizing communities.

In Fig.~\ref{Fig:Kuramoto}A, we depict the block-matrix community structure in a
network of $128$ Kuramoto oscillators with $8$ equally-sized communities. The
phase $\theta_{i}(t)$ of the $i^{\mathrm{th}}$ oscillator evolves in time
according to
\begin{equation}
	\frac{d \theta_{i}}{dt} = \omega_{i} + \sum_{j} \kappa C_{ij} \mathrm{sin}(\theta_{j}-\theta_{i})\,, \quad i \in \{1, \ldots, N\}\,,
\label{kuramoto}
\end{equation}
where $\omega_{i}$ is the natural frequency of oscillator $i$, the matrix
$\mathbf{C}$ gives the binary-valued ($0$ or $1$) coupling between each pair
of oscillators, and $\kappa$ (which we set to $0.2$) is a positive real
constant that indicates the strength of the coupling. We draw the frequencies
$\omega_i$ from a Gaussian distribution with mean $0$ and standard deviation
$1$. Each node is connected to 13 other nodes (chosen uniformly at random) in
its own community and to one node outside of its community.  This external
node is chosen uniformly at random from the set of all nodes from other
communities.

To quantify the temporal evolution of synchronization patterns, we define a
set of temporal networks from the time-dependent correlations (which,
following Ref.~\cite{Arenas2006}, we use to measure synchrony) between pairs
of oscillators: $A_{ij}(t) = \left\langle |\cos[\theta_{i}(t) -
\theta_{j}(t)]| \right\rangle$, where the angular brackets indicate an
average over $20$ simulations. We perform simulations, each of which use a
different realization of the coupling matrix $\mathbf{C}$ (see the
Supplementary Material [24] for details of the numerics). Importantly, edge
weights not only vary (see Fig.~\ref{Fig:Kuramoto}B) but they also
\emph{co-vary} with one another (see Fig.~\ref{Fig:Kuramoto}C) in time: the
strength of network co-evolution, which we denote by $\nu_{A_{t}}$, is
greater than that expected in a null-model network in which each edge-weight
time series is independently permuted uniformly at random.

In this example, the cross-links given by the non-zero elements of ${\bf
\Lambda^{\prime}}$ form a single connected component due to the extensive
co-variation. One can distinguish cross-links according to their roles
relative to the community structure in Fig.~\ref{Fig:Kuramoto}A
\cite{Guimera2005}: (i) pairs of within-community edges, (ii) pairs of
between-community edges, and (iii) pairs composed of one within-community
edge and one between-community edge. Assortative pairings [i.e., cases (i)
and (ii)] are significantly more represented than disassortative pairings
[i.e., case (iii)] (see Fig.~\ref{Fig:Kuramoto}D). The assortative nature of
cross-links might be driven by the underlying community structure in the
block structure in Fig.~\ref{Fig:Kuramoto}A: within-community edges are
directly connected to one another via shared nodes, whereas between-community
edges are more distantly connected to one another via a common input (e.g., a
sparse but frequently updating representation of the state of other
oscillators).

Using community detection, we identified 5 distinct edge sets (i.e.,
\emph{hyperedges}) in ${\bf \Lambda^{\prime}}$ with distinct temporal
profiles (see Fig.~\ref{Fig:Kuramoto}E). The first hyperedge tends to connect
within-community edges to each other. On average, they tend to synchronize
early in our simulations. The second and third hyperedges tend to connect
between-community edges to each other.  The second hyperedge connects edges
that tend to exhibit a late synchronization, and the third one connects edges
that tend to exhibit an initial synchronization followed by a
desynchronization. The fourth and fifth hyperedges are smaller in size (i.e.,
contain fewer edges) than the first three, and their constituent edges
oscillate between regimes with high and low synchrony. The edges that
constitute the fifth hyperedge oscillate at approximately one frequency,
whereas those in the fourth hyperedge have multiple frequency components. See
the Supplementary Material [24] for a characterization of the temporal
profiles and final synchronization patterns of hyperedges in the network of
Kuramoto oscillators.

Together, our results demonstrate the presence of multiple co-evolution
profiles: early synchronization, late synchronization, desynchronization, and
oscillatory behavior \cite{Arenas2008}. Moreover, the assortative pairing of
cross-links indicates that temporal information in this dynamic system is
segregated not just within separate synchronizing communities but also in
between-community edges.


\section{Networks of Human Brain Areas}

Our empirical data captures the changes in regional brain activity over time
as experimental subjects learn a complex motor-sequencing task that is
analogous to playing complex keyboard arpeggios. Twenty individuals practiced
on a daily basis for 6 weeks, and we acquired MRI brain scans of blood
oxygenated-level-dependent (BOLD) signal at four times during this period. We
extracted time series of MRI signals from $N=112$ parts of each individual's
brain \cite{Bassett2013}. Co-variation in BOLD measurements between brain
areas can indicate shared information processing, communication, or input;
and changes in levels of coherence over time can reflect the network
structure of skill learning. We summarize such functional connectivity
\cite{Friston1994} patterns using an $N \times N$ coherence matrix
\cite{Bassett2011b,Bassett2012b}, which we calculate for each experimental
block. We extract temporal networks, which each consist of 30 time points,
for naive (experimental blocks corresponding to 0--50 trials practiced),
early (60--230), middle (150--500), and late (690--2120) learning
\cite{Bassett2013}. We hypothesize that learning should be reflected in
changes of hypergraph properties over the very long time scales (6 weeks)
associated with this experiment.

\begin{figure}
\begin{center}
\includegraphics[width=0.95\linewidth]{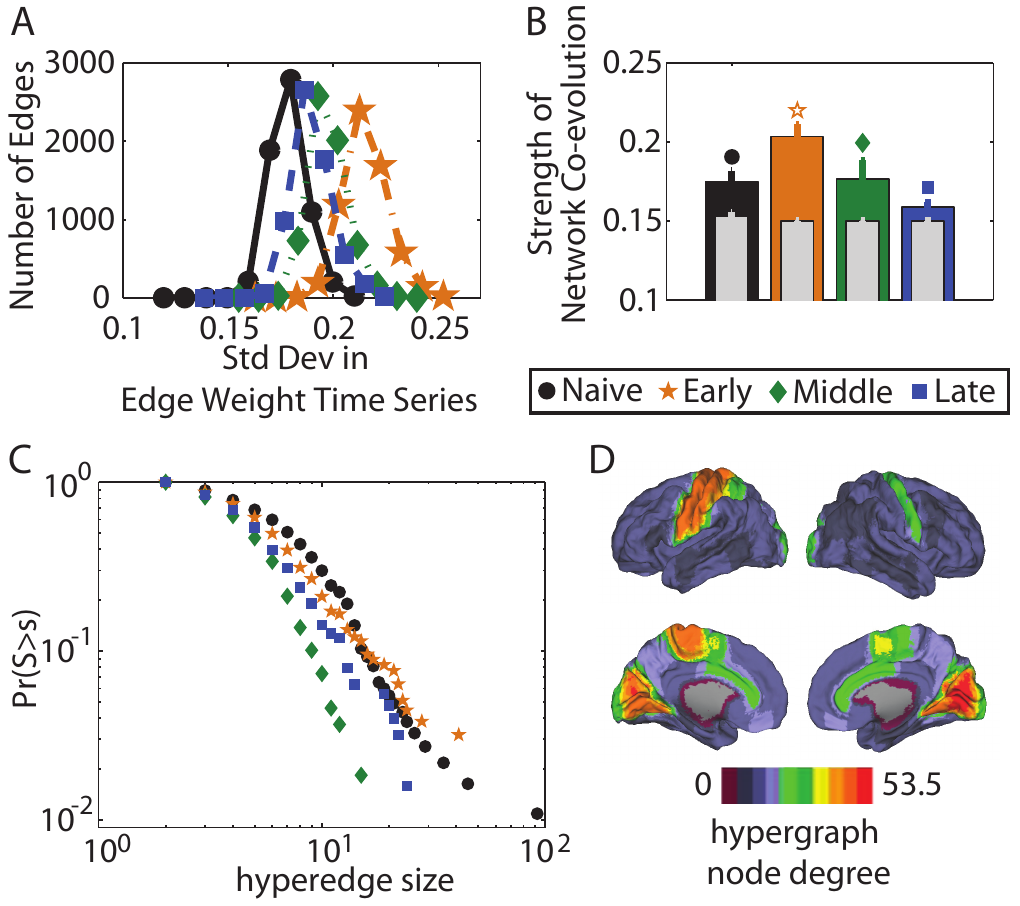}
\caption[]{\label{fig:hyper1} \textbf{Co-evolution Properties of Brain
Network Dynamics.} \emph{(A)} A histogram of the number of edges as a
function of the standard deviation in edge weights over time for the 4
temporal networks. \emph{(B)} Strength of network co-evolution $\nu_{A_{t}}$
of 4 temporal networks and the respective null-model networks (gray). Error
bars indicate standard deviation of the mean over study participants.
\emph{(C)} Cumulative probability distribution $Pr$ of the size $s$ of
hyperedges in the 4 learning hypergraphs. \emph{(D)} Anatomical distribution
of early-learning hypergraph node degree (averaged over the 20 participants).
We obtain qualitatively similar results from the early, middle, late, and
extended learning temporal networks. In panels \emph{(A-C)}, color and shape
indicate the temporal network corresponding to (black circles) naive, (orange
stars) early, (green diamonds) middle, and (blue squares) late learning. }
    \end{center}
\end{figure}

Temporal brain networks exhibit interesting dynamics: all four temporal
networks exhibit a non-zero variation in edge weights over time (see
Fig.~\ref{fig:hyper1}A). Importantly, edge weights not only vary but
\emph{co-vary} in time: the strength of network co-evolution $\nu_{A_{t}}$ is
greater in the 4 real temporal networks than expected in a random null-model
network in which each edge-weight time series is independently permuted
uniformly at random (see Fig.~\ref{fig:hyper1}B). The magnitude of temporal
co-variation between functional connections is modulated by learning: it is
smallest prior to learning and largest during early learning (i.e., amidst
most performance gains). These results are consistent with the hypothesis
that the adjustment of synaptic weights during learning alters the
synchronization properties of neurophysiological signals \cite{Bassett2011b},
which could manifest as a steep gain in the co-evolution of synchronized
activity of large-scale brain areas.

To uncover groups of co-evolving edges, we study the edge-edge correlation
matrix ${\bf \Lambda^{\prime}}$, whose density across the 4 temporal networks
and the 20 study participants ranged from approximately 1\% to approximately
95\%. We found that the significant edges were already associated with
multiple connected components, so we did not further partition the edge sets
into communities. The distribution of component sizes $s$ is heavy-tailed
(see Fig.~\ref{fig:hyper1}C), which perhaps reflects inherent variation in
the communication patterns that are necessary to perform multiple functions
required during learning \cite{Bassett2011b}. With long-term training,
hyperedges decrease in size (see Fig.~\ref{fig:hyper1}C), which might reflect
an emerging autonomy of sensorimotor regions that can support sequential
motor behavior without relying on association cortex.

Hyperedges indicate temporal co-variation of putative communication routes in
the brain and can be distributed across different anatomical locations. The
hypergraph node degree quantifies the number of hyperedges that are connected
to each brain region. We observe that nodes with high hypergraph degree are
located predominantly in brain regions known to be recruited in motor
sequence learning \cite{Dayan2011}: the primary sensorimotor strip in
superior cortex and the early visual areas located in occipital cortex (see
Fig.~\ref{fig:hyper1}D).


\section{Methodological Considerations \& Future Directions}

The approach that we have proposed in this paper raises several interesting
methodological questions that are worth additional study.

First, there are several ways (e.g., using the edge-edge correlation matrix
${\bf \Lambda}$) to define the statistical significance of a single element
in a large matrix that is constructed from correlations or other types of
statistical similarities between time series (see the Supplementary Material
[24]).  Naturally, one should not expect that there is a single
``best-choice'' correction for false-positive (i.e., Type I) errors in these
matrices that is applicable to all systems, scales, and types of association.
In the future, rather than using a single threshold for statistical
significance to convert ${\bf \Lambda}$ to ${\bf \Lambda^{\prime}}$, it might
be advantageous to use a range of thresholds --- perhaps to differentially
probe strong and weak elements of a correlation matrix, as has been done in
the neuroimaging literature \cite{Bassett2012a} --- to characterize the
organization of the hypergraphs on different geometrical scales (i.e., for
different distributions of edge-weight values).

Second, the dependence of the hypergraph structure on the amount of time $T$
that we consider is also a very interesting and worthwhile question.
Intuitively, the hypergraph structure seems to capture transient dependencies
between edges for small $T$ but to capture persistent dependencies between
edges for large $T$. A detailed probing of the $T$-dependence of the
hypergraph structure could be particularly useful for studying systems that
exhibit (i) temporally-independent state transitions based on their
cross-linked structures and (ii) co-evolution dynamics that occur over
multiple temporal scales.

Finally, the approach that we have proposed in this paper uses hypergraphs to
connect dependencies between interactions to the components that interact.
Alternatively, one can construe the interactions themselves as one network
and the components that interact as a second network.  This yields a
so-called \emph{interconnected network} (which is a type of \emph{multilayer
network} \cite{Kivela2013}), and the development of techniques to study such
networks is a burgeoning area of research.  Using this lens makes it clear
that our approach can also be applied ``in the other direction" to connect
sets of components that exhibit similar dynamics (one network) to
interactions between those components (another network). This yields a simple
multilayer structure in which a single set of components is connected by two
sets of associations (similarities in dynamics and via a second type of
interaction).  However, we believe that the ``forward" direction that we have
pursued is the more difficult of the two directions, as one needs to connect
a pair of networks whose edges are defined differently and whose nodes are
also defined differently. Hypergraphs provide one solution to this difficulty
because they make it possible to bridge these two networks. Moreover, many
dynamical systems include both types of networks: a network that codifies
dependencies between nodes and a network that codifies dependencies between
node-node interactions.


\section*{Conclusion}

Networked systems are ubiquitous in technology, biology, physics, and
culture. The development of conceptual frameworks and mathematical tools to
uncover meaningful structure in network dynamics is critical for the
determination and control of system function. We have demonstrated that the
cross-link structure of network co-evolution, which can be represented
parsimoniously using hypergraphs, can be used to identify unexpected temporal
attributes in both real and simulated temporal dynamical systems. This, in
turn, illustrates the utility of analyzing cross-links for investigating the
structure of temporal networks.


\section*{Acknowledgements}

We thank Aaron Clauset for useful comments. We acknowledge support from the
Sage Center for the Study of the Mind (DSB), Errett Fisher Foundation (DSB),
James S. McDonnell Foundation (\#220020177; MAP), the FET-Proactive project
PLEXMATH (FP7-ICT-2011-8, Grant \#317614; MAP) funded by the European
Commission, EPSRC (EP/J001759/1; MAP), NIGMS (R21GM099493; PJM), PHS
(NS44393; STG), and U.S. Army Research Office (W911NF-09-0001; STG). The
content is solely the responsibility of the authors and does not necessarily
represent the official views of any of the funding agencies.

\bibliographystyle{apsrev4-1} 
\bibliography{bibfile5}


\end{document}


\title{Supplemental Material for \\ ``Cross-Linked Structure of Network Evolution"}

\author{Danielle S. Bassett$^{1,2,3,*}$, Nicholas F.
Wymbs$^{4}$, Mason A. Porter$^{5,6}$, Peter J. Mucha$^{7,8}$, Scott T.
Grafton$^{4}$} \affiliation{$^{1}$Department of Bioengineering, University of
Pennsylvania, Philadelphia, PA 19104, USA;\\ $^{2}$Department of Physics,
University of California, Santa Barbara, CA 93106, USA;\\ $^{3}$ Sage Center
for the Study of the Mind, University of California, Santa Barbara, CA
93106;\\ $^{4}$
Department of Psychology and UCSB Brain Imaging Center, University of California, Santa Barbara, CA 93106, USA;\\
$^{5}$ Oxford Centre for Industrial and Applied Mathematics, Mathematical
Institute, University of Oxford, Oxford OX1 3LB, UK;\\ $^{6}$ CABDyN
Complexity Centre, University of Oxford, Oxford, OX1 1HP, UK;\\
$^{7}$Carolina Center for Interdisciplinary Applied Mathematics, Department
of Mathematics, University of North Carolina, Chapel Hill, NC 27599, USA;\\
$^{8}$Department of Applied Physical Sciences, University of North Carolina, Chapel Hill, NC 27599, USA;\\
$^*$Corresponding author. Email address: dsb@seas.upenn.edu}

\date{\today}

\maketitle

\newpage

In this supplementary document, we include the following material to support
the work described in the main text.
\begin{enumerate}
    \item A detailed description of statistical corrections for edge-edge
        association matrices.
    \item A description of the community-detection techniques that we applied to
        the edge-edge association matrix.
    \item A characterization of the temporal profiles and final
        synchronization patterns of hyperedges in the network of Kuramoto
        oscillators.
    \item A comparison to null models based on surrogate data.
    \item A note on numerical implementation.
    \item Figure S1: Hyperedge Identification in a Network of Kuramoto
        Oscillators.
    \item Figure S2: Final Synchronization Patterns and Temporal Profiles
        of Hyperedges.
\end{enumerate}

\newpage


\section{Statistical Corrections for Edge-Edge Association Matrices}

In the main text, we describe a method for extracting cross-links from
temporal networks. For a temporal network $\bf{A_{t}}$, we calculate the $E
\times E$ adjacency matrix ${\bf \Lambda}$, where the matrix element
$\Lambda_{ab}$ gives the Pearson correlation coefficient between the time
series of weights for edge $a$ and the time series of weights for edge $b$.
Note that $E=N(N-1)/2$ is the total number of possible (undirected) edges per
layer in a temporal network.  (Each layer can come from a single point in
time, aggregation over a given time window, etc.) For simplicity, we employ a
correlation coefficient as a measure of statistical association to examine
linear relationships in ensembles of edge-weight time series
\footnote{However, the methods developed in this paper also work for other
choices of statistical association between edge-weight time series.}. Because
we seek to determine sets of edges that might have a common driver, we do not
employ sparse network methods such as the graphical lasso \cite{Friedman2008}
or Bayesian network \cite{Hinton1997} methods that attempt to estimate
pairwise relationships between time series in a manner that is independent of
other variables.

Given the very large number of statistical tests that the above procedure
entails, we threshold the edge-edge correlation matrix ${\bf \Lambda}$ to
retain only statistically significant connections, which we determine by
estimating the p-value associated with the Pearson coefficient $r$ for each
edge-edge correlation. Using a false-positive correction for multiple
comparisons, we threshold ${\bf \Lambda}$ by identifying significant matrix
elements as those whose associated p-value satisfies
\begin{equation} \label{pval_fpc}
	p<\frac{1}{M}=\frac{2}{E(E-1)}\,,
\end{equation}
where $M$ is the number of tests that were performed. We retain the original
weights of significant matrix elements in a new matrix ${\bf
\Lambda^{\prime}}$ and set nonsignificant matrix elements to $0$.

The type of multiple comparisons correction that one uses to control for Type
I errors (i.e., false positives) in correlation matrices derived from (both
real and simulated) dynamical systems is itself interesting
\cite{Prignano2012,Fornito2013}. The false-positive correction of $p<1/M$
that we applied is an increasingly common choice in the study of correlation
matrices in the neuroscience literature
\cite{AlexanderBloch2010,Messe2013,Bassett2009,Lynall2010,Weiss2011,Jin2011,Liao2013}.
It has been argued that alternative choices, such as the false-discovery rate
\cite{Benjamini1995,Benjamini2001} and Bonferroni-correction methods
\cite{Hochberg1988,Dunn1961,Dunnett1955}, are too stringent for situations
like correlation matrices in which variables are highly inter-dependent
\cite{Fornito2013}, and they can lead to an overly large number of Type II
errors (i.e., false negatives) \cite{Roxy2011}.

After performing the statistical correction to obtain the weighted
thresholded matrix ${\bf \Lambda^{\prime}}$, we wish to extract cohesive sets
of co-evolving edges. Two potential cases are apparent. The simpler case
occurs when ${\bf \Lambda^{\prime}}$ is composed of disconnected components
that each contain a set of co-evolving edges. We illustrate this scenario in
the main manuscript using networks of brain regions. In a second case, ${\bf
\Lambda^{\prime}}$ contains a single large connected component --- which can
but need not include all of a network's edges --- from which one must further
extract sets of co-evolving edges.  We illustrate this scenario, which arises
from extensive and broadly distributed temporal covariance, in the main
manuscript using networks of Kuramoto oscillators.

To study the second scenario, we need to use a method for extracting sets of
strongly cross-linked edges in ${\bf \Lambda^{\prime}}$. One possible
approach is to choose a more stringent statistical threshold for creating
${\bf \Lambda^{\prime}}$ in the first place.  For example, one could tune the
threshold so that it fragments ${\bf \Lambda}$ into several disconnected
components. However, such an approach requires the choice of an arbitrarily
stringent threshold on the p-value $p$ and entails the risk of Type II errors
(i.e., false negatives) \cite{Fornito2013}.  In this paper, we employ an
alternative approach: we extract sets of strongly cross-linked edges using
community detection \cite{Porter2009,Fortunato2010}. An advantage of this
approach is that we can exploit the complete information housed in ${\bf
\Lambda^{\prime}}$ by using community-detection methods that account for
cross-link weights and their signs \cite{Traag2009}.


\section{Community Detection on Edge-Edge Association Matrices}

Methods for detecting communities in networks make it possible to
algorithmically extract groups of nodes that are highly and mutually
interconnected \cite{Porter2009,Fortunato2010,Newman2012}. In this paper, we
seek sets of edges that are strongly and densely cross-linked to one
another\footnote{It is important to note that the identification of
communities of edges based on the temporal covariance of their edge-weight
time series is different from the identification of ``edge communities'',
which consist of set of edges in static undirected networks that share common
nodes \cite{Ahn2010}.}.  We identify such ``communities" (or ``modules") of
edges by optimizing a \emph{modularity quality function} that is suitable for
signed matrices \cite{Traag2009}:
\begin{equation}\label{eq:one}
	Q = \sum_{ab} \left[\Lambda^{\prime}_{ab} -\gamma^{+} P_{ab}^{+} + \gamma^{-} P_{ab}^{-}\right] \delta(g_{a},g_{b})\,,
\end{equation}
where ${\bf \Lambda^{\prime}} = (\Lambda^{\prime}_{ab})$ is the $E \times E$
thresholded and weighted correlation matrix, edge $a$ is assigned to
community $g_{a}$, edge $b$ is assigned to community $g_{b}$, the Kronecker
delta $\delta(g_{a},g_{b})=1$ if $g_{a} = g_{b}$ and it equals $0$ otherwise,
$\gamma^{+}$ and $\gamma^{-}$ are resolution parameters, and $P_{ab}^{+}$ and
$P_{ab}^{-}$ are the respective expected weights of the positive and negative
cross-links that connect edge $a$ and edge $b$ via a specified null model. We
employ a signed null model \cite{Traag2009} with $\gamma^{+}=\gamma^{-}$
(also see \cite{Gomez2009}, who study the case $\gamma^{+}=\gamma^{-}=1$), so
that
\begin{equation}
	P_{ab}^{+} = \frac{k^{+}_{a} k^{+}_{b}}{\sum_{ab} k^{+}}\,,\qquad
	P_{ab}^{-} = \frac{k^{-}_{a} k^{-}_{b}}{\sum_{ab} k^{-}}\,,\\
\end{equation}
where $k_{a}^{\pm}=\sum_b \Lambda^{\prime \pm}_{ab}$ is the strength of
cross-link $a$ in the matrix ${\bf \Lambda^{\prime \pm}}$.  The matrix ${\bf \Lambda^{\prime +}}$ retains all positively weighted elements of $\Lambda^{\prime}_{ab}$ and sets
all negatively weighted elements of $\Lambda^{\prime}_{ab}$ to
$0$. The matrix ${\bf \Lambda^{\prime -}}$ retains all negatively weighted elements of $\Lambda^{\prime}_{ab}$ and sets
all positively weighted elements of $\Lambda^{\prime}_{ab}$ to
$0$.

Maximization of $Q$ yields a hard partition of the edge-edge network into
communities such that the total cross-link weight inside of communities is as
large as possible (relative to the null model and subject to the limitations
of the employed computational heuristics, as optimizing $Q$ is NP-hard
\cite{Porter2009,Fortunato2010,Brandes2008}). Given the near-degeneracy of
the landscape of the modularity function $Q$ \cite{Good2010}, we perform 100
optimizations of Eq.~2 and obtain consensus partitions over these
optimizations via a comparison to an appropriate null model. (See
Ref.~\cite{Bassett2012b} for a detailed description of the method.)

The structural resolution parameter $\gamma = \gamma^{+}=\gamma^{-}$ is a
tunable scalar that sets the size of the communities in the (near) optimal
partition \cite{Lancichinetti2011,Bassett2012b}. Small values of $\gamma$
produce a few large communities, whereas large values of $\gamma$ produce
many small communities. By tuning $\gamma$, one can therefore examine the
community structure at different scales
\cite{Onnela2012,Kumpula2007,Lambiotte2008,Delvenne2010,Schaub2012b,Schaub2012}
of both real
\cite{Meunier2009,Meunier2010,Bassett2010,Bassett2010c,Bassett2011b} and
simulated \cite{Arenas2006,Arenas2006b} dynamical systems.

\begin{figure}
\begin{center}
\includegraphics[width=0.7\linewidth]{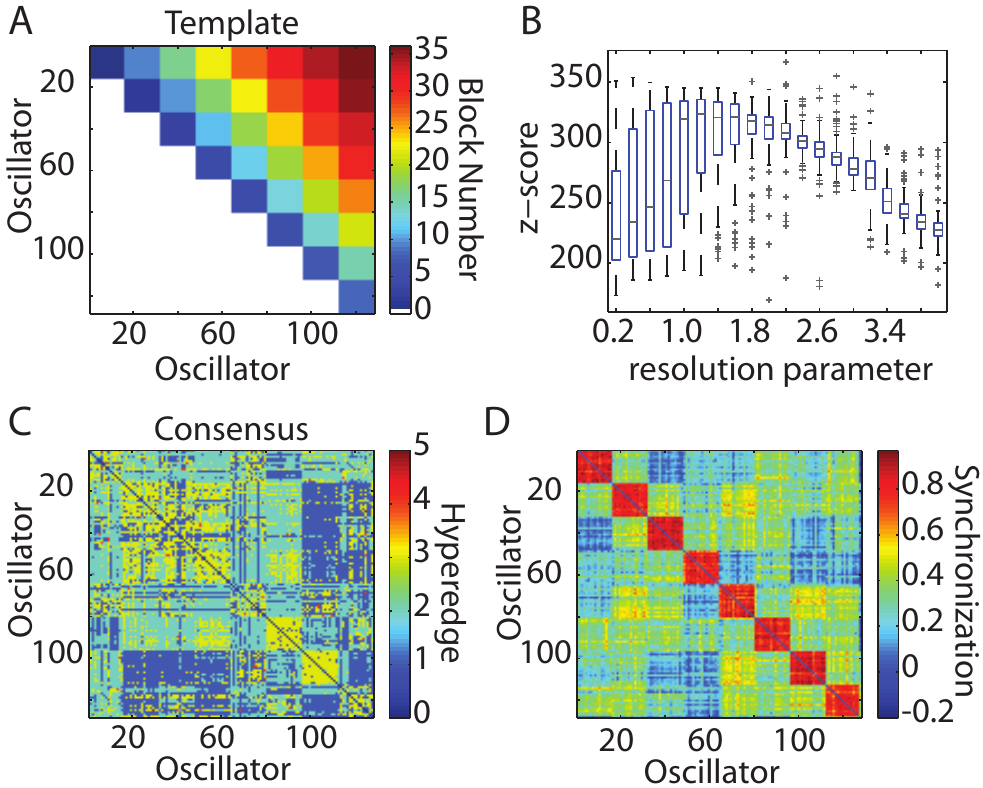}
\caption[]{\label{fig:gamma} \textbf{Hyperedge Identification in a Network of
Kuramoto Oscillators}. \emph{(A)} Template indicating the block structure of
the community structure of nodes in a network of 128 Kuramoto oscillators.
Blocks 1--8 contain within-community edges, and blocks 9--36 contain
between-community edges. Color indicates the block number. \emph{(B)} The
z-score of the Rand coefficient between the upper triangle of the template in
\emph{(A)} and the partition of the thresholded and weighted edge-edge
correlation matrix ${\bf \Lambda^{\prime}}$ into communities of edges. Box
plots indicate quartiles and 95\% confidence intervals over the 100
optimizations of the signed modularity quality function in Eq.~2.
\emph{(C)} Consensus over partitions obtained from 100 optimizations at
$\gamma=1.8$. Each community of edges constitutes a \emph{hyperedge}, and
color indicates hyperedge number. \emph{(D)} The final synchronization
pattern of the network of Kuramoto oscillators at the final time ($T=100$),
which is reminiscent of the community structure of the network (which we show
in Fig.2A in the main manuscript). Color indicates time-dependent correlation
between pairs of oscillators (which we use to indicate their level of
synchrony, following \cite{Arenas2006}): $A_{ij}(t) = \left\langle
|\cos[\theta_{i}(t) - \theta_{j}(t)]| \right\rangle$, where the angular
brackets indicate an average over $20$ simulations.
    }
    \end{center}
\end{figure}

For simplicity, we focus on a single resolution-parameter value for detailed
investigation. We choose a value that provides insight into the relationship
between the community structure of edges and the community structure of
nodes, and later we discuss at length the procedure that we used to select
this value. In Fig.~\ref{fig:gamma}A, we show a template block structure that
summarizes the community structure of nodes in a network of Kuramoto
oscillators. Each block contains edges that are located either (i) within
communities (template blocks 1--8) or (ii) between communities (template
blocks 9--36). We characterize the similarity between this template (which
yields a network partition that we label by $\alpha$) and the community
structure of edges at a given value of the structural resolution parameter
$\gamma$ (which yields a partition that we label by $\beta$) using the
z-score of the Rand coefficient \cite{Traud2010}. We use $w_{11}$ to denote
the count of edge pairs that are classified together in both partitions
(e.g., $\alpha$ and $\beta$). We use $w_{10}$ to denote the count of edge
pairs that are classified together in the first partition but classified
separately in the second partition, and we define $w_{01}$ analogously as the
count of edge pairs that are classified separately in the first partition but
classified together in the second partition.  We use $w_{00}$ to denote the
count of edge pairs that are classified separately in both partitions. The
total number $R$ of node pairs is then given by the sum of these quantities:
$R = w_{11}+w_{10}+w_{01}+w_{00}$. We calculate the Rand $z$-score in terms
of the network's total number of node pairs $R$, the number of pairs
$R_{\alpha}$ classified the same way in partition $\alpha$, the number of
pairs $R_{\beta}$ classified the same way in partition $\beta$, and the
number of node pairs $w_{\alpha \beta}$ that are assigned to the same
community both in partition $\alpha$ and in partition $\beta$. The $z$-score
of the Rand coefficient comparing these two partitions is
\begin{equation}
	z_{\alpha\beta} = \frac{1}{\sigma_{w_{\alpha \beta}}}
	\left(w_{\alpha \beta}-\frac{R_{\alpha}R_{\beta}}{R}\right)\,,
\end{equation}
where $\sigma_{w_{\alpha \beta}}$ is the standard deviation of $w_{\alpha
\beta}$ (as in \cite{Traud2010}).

In the resolution parameter range $\gamma \in [0.2,4]$, the z-score of the
Rand coefficient between the template and partitions into communities of
edges appears to have two regimes (see Fig.~\ref{fig:gamma}B). For $\gamma
\lessapprox1.8$, the z-score exhibits are large variability over multiple
optimizations of the modularity quality function in Eq.~2, which suggests
that the optimization landscape of $Q$ is replete with local maxima
\cite{Good2010}. However, for $\gamma \gtrapprox 1.8$, the z-score has a much
smaller variability over the multiple optimizations, which suggests that the
partitions in this regime are relatively robust \cite{Bassett2012b}. In this
second regime, ($\gamma \gtrapprox 1.8$), the mean z-score also decreases
with increasing $\gamma$, which indicates that partitions with a large number
of small communities (i.e., for $\gamma$ values closer to $4$) exhibit less
similarity to the template than partitions with a small number of large
communities (i.e., $\gamma$ values closer to $1.8$).

We choose to examine the community structure in the edge-edge correlation
matrix at the resolution parameter $\gamma=1.8$ for two reasons: (i) at this
resolution-parameter value, partitions are more robust (i.e., less variable)
over multiple optimizations than they are at lower values of $\gamma$; and
(ii) this approximately maximizes the similarity, as measured by the Rand
z-score, between the community structure of edges and the community structure
of nodes. To visualize the cross-linked edge communities that are present at
$\gamma=1.8$, we construct a consensus partition \cite{Lancichinetti2012}
over the 100 optimizations using a method that corrects for statistical noise
in sets of partitions defined in comparison to a null model
\cite{Bassett2012b}. The consensus partition assigns each edge to one of 5
communities of varying sizes (see Fig.~\ref{fig:gamma}C). Each community
yields a hyperedge, and we note that the pattern of hyperedges in the network
has an inherently different structure than the final synchronization pattern
of the network of Kuramoto oscillators (compare Figs.~\ref{fig:gamma}C and D)
\footnote{Note that we adopt the notion of synchronization from
Ref.~\cite{Arenas2006} and measure synchrony between a pair of oscillators in
terms of the time-dependent correlation between them.}. In the next section,
we characterize the differences between these two structures in greater
detail.


\section{Temporal Profiles and Final Synchronization Patterns of Hyperedges in the Network of Kuramoto Oscillators}

\begin{figure}
\begin{center}
\includegraphics[width=1\linewidth]{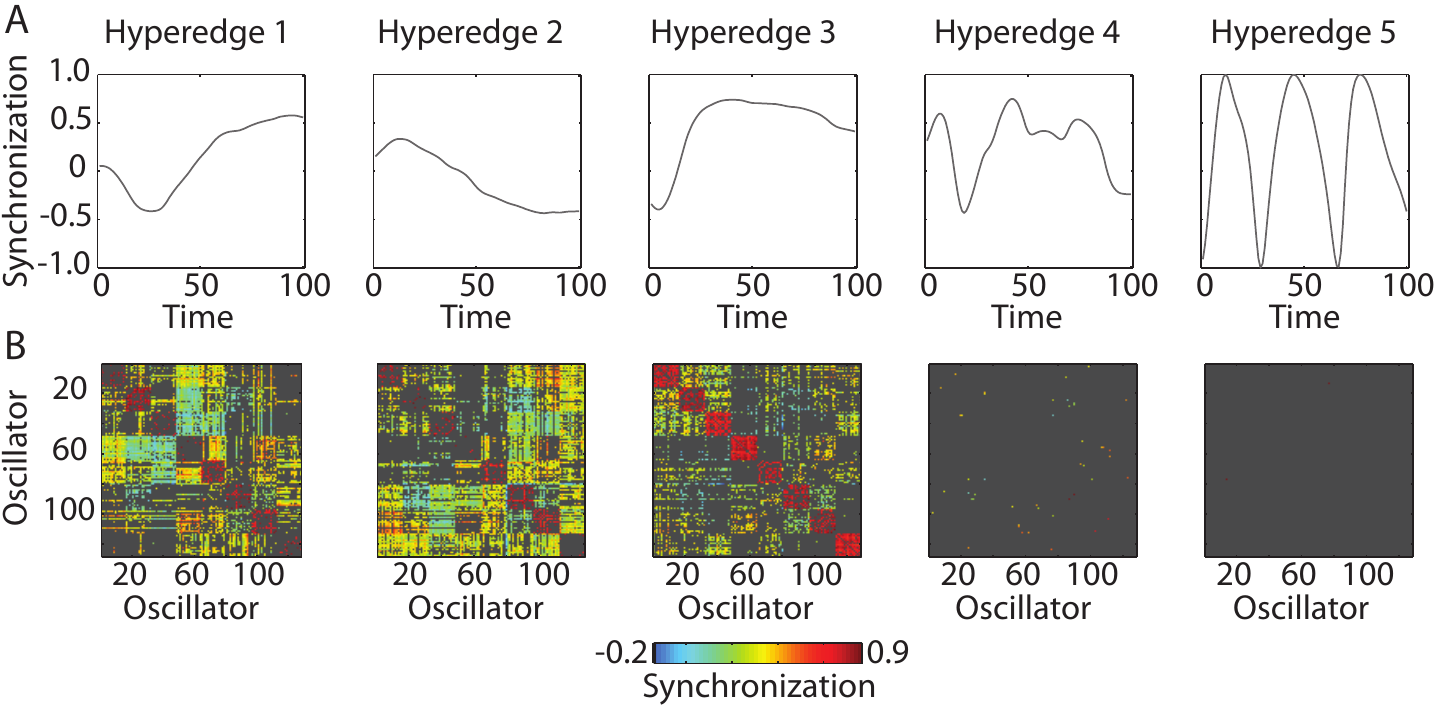}
\caption[]{\label{fig:synch} \textbf{Temporal Profiles and Final
Synchronization Patterns of Hyperedges}. \emph{(A)} The mean synchronization
of edges as a function of time [$S(t) = \sum_{(i,j) \in h} A_{ij}(t)$] and
\emph{(B)} the final synchronization weights of each edge.  From left to
right, we plot these for hyperedge 1 (in the left panel) to hyperedge 5
(right panel). Color indicates time-dependent correlation between pairs of
oscillators: $A_{ij}(t) = \left\langle |\cos[\theta_{i}(t) - \theta_{j}(t)]|
\right\rangle$, where the angular brackets indicate an average over $20$
simulations. Matrix elements highlighted in gray indicate edges that are
members of a hyperedge other than their own.
    }
    \end{center}
\end{figure}

Each hyperedge that we identified in the network of Kuramoto oscillators
consists of a set of edges with a different temporal weight profile (see
Fig.~\ref{fig:synch}A). Edges are cross-linked based on the similarity in
their temporal weight profile, and community detection makes it possible to
extract cohesive groups of edges with similar profiles. The first two
hyperedges, whose dynamics we show in the left two panels of
Fig.~\ref{fig:synch}, tend to consist of between-community edges (see
Fig.~\ref{fig:synch}B) and exhibit either late increases in weight (which
yields late synchronizaton) or decreases in weight (desynchronization) over
time (see Fig.~\ref{fig:synch}A). The hyperedge whose dynamics we show in the
center panel of Fig.~\ref{fig:synch} includes the majority of the
within-community edges and exhibits a strong increase in weight (and hence
oscillator synchronization) early in the simulation. The final two
hyperedges, whose dynamics we show in the right two panels of
Fig.~\ref{fig:synch}, consist of edges that exhibit high-frequency
oscillatory behavior in their weights.

Our investigation of cross-links and subsequent hyperedge extraction
identifies similarities between edges that are based on their temporal
profiles and can therefore be different from their final synchronization
values. For example, hyperedges 1--3 in Fig.~\ref{fig:synch} each include
edges with a wide range of final synchronization values that range from very
strong ($A_{ij} \doteq 0.9$) to very weak ($A_{ij} \doteq -0.2$). Each
hyperedge instead captures a property of edge dynamics: the trajectory that
that edge followed to attain a given final synchronization value.

\section{Comparison to Null Models Based on Surrogate Data}

When examining networks that are extracted from real data, it is important to determine when observed structures different significantly from those in a relevant null-model
system \cite{Bassett2012b}. Specifically, for networks constructed from
statistical similarities between time series (such as the brain networks
that we examine in this paper), one can construct null models based on surrogate
time series. By comparing cross-link structure in the real and null-model
systems, one can probe potentially meaningful features of co-evolution in the real network.

We employ a surrogate-data generation method that has been used previously to construct
covariance matrices \cite{Nakatani2003} and to characterize static
\cite{Zalesky2012} and temporal \cite{Bassett2012b} networks. The
\emph{Fourier transform (FT) surrogate} scrambles the phase of time series in
Fourier space \cite{Prichard1994} and thereby preserves the mean, variance, and
autocorrelation function of the original time series. We assume that the
linear properties of the time series are specified by the squared amplitudes
of the discrete Fourier transform
\begin{equation}
	|S(u)|^2 = \left| \frac{1}{\sqrt{V}} \sum_{v=0}^{V-1} s_{v} e^{i 2 \pi u v / V}\right|^{2}\,,
\end{equation}
where $s_{v}$ denotes an element in a time series of length $V$. (That is,
$V$ is the number of elements in the time-series vector.)  We construct
surrogate data by multiplying the Fourier transform by phases chosen
uniformly at random and then transforming back to the time domain:
\begin{equation}
	\bar{s}_{v} = \frac{1}{\sqrt{V}} \sum_{v=0}^{V-1} e^{i a_{u}} |S_{u}| e^{i 2 \pi k v / V}\,,
\end{equation}
where $a_{u}\in [0,2 \pi)$ are chosen independently and uniformly at random
\footnote{The code that we used for this computation actually operates on
$[0,2 \pi]$, which is the same as $[0,2 \pi)$ except for a set of measure 0.}

We construct FT surrogate time series from the original time series that we
extracted from each brain region of each subject during each scanning
session. Using identical procedures to those that we employed to study the
real time series, we cut each surrogate time series into time windows that
correspond to trial blocks, compute the coherence between pairs of surrogate
time series, calculate the thresholded edge-by-edge correlation matrix ${\bf
\Lambda^{\prime}}$, and extracted hyperedges defined as the connected
components ${\bf \Lambda^{\prime}}$. In contrast to the heavy-tailed
hyperedge-size distributions that we observe in the real data (see Fig.~3C of
the main manuscript), we find that the size distributions extracted from the
surrogate data are narrow and peaked: $s \approx 2.09\pm0.29$ (mean $\pm$
standard deviation) for naive learning, $s \approx 2.09\pm0.29$ for early
learning, $s \approx 2.13\pm0.34$ for middle learning, and $s \approx
2.20\pm0.40$ for late learning. The maximum hyperedge size is $3$ and the
minimum is $2$. These results demonstrate that the learning-related human
brain co-evolution structure that we report in the main manuscript cannot be
attributed to the mean, variance, or autocorrelation function of the original
time series.


\section{A Note on Numerical Simulation}

To simulate the dynamics of the network of Kuramoto oscillators, we solve
the discrete-time equation
\begin{equation}\label{kur}
	\theta_{t} = \theta_{t-1}+\tau \omega_{i} + \sum_{j} \kappa C_{ij}
\mathrm{sin}(\theta_{j}-\theta_{i})\,,
\end{equation}
where $\omega_{i}$ is the natural frequency of oscillator $i$, the matrix
$\mathbf{C}$ gives the binary-valued ($0$ or $1$) coupling between each pair
of oscillators, $\tau$ (which we set to 0.1) is a positive real constant that
indicates the time step, and $\kappa$ (which we set to $0.2$) is a positive
real constant that indicates the strength of the coupling. We solve equation (\ref{kur}) for $t \in \{1, \ldots, T\}$ for a maximum of $T=101$ time points. We base our simulation method on the implementation in Ref~\cite{Sumpter2013}. Each matrix in the temporal network $\mathbf{A}_{t}$
gives the time-dependent correlations, measured at time point $t$, between pairs of oscillators.

\clearpage
\newpage

\bibliographystyle{apsrev4-1} 
\bibliography{bibfile3}
